\documentclass[doublecol]{epl2} 

\RequirePackage{fancyhdr}

\RequirePackage{xcolor}

\RequirePackage{graphicx}

\RequirePackage{amssymb}

\RequirePackage{amsmath}

\RequirePackage{cuted}

\title{Thermodynamic Geometry of two-dimensional square-well fluids}

\author{Jaime Jaramillo-Gutiérrez\inst{1} \and José Torres-Arenas\inst{1,2}}
\shortauthor{Jaramillo-Gutiérrez \etal}

\institute{                    
  \inst{1} División de Ciencias e Ingenierías, Campus León, Universidad de Guanajuato, México.\\
  \inst{2} Departamento de Física Aplicada, Universidade de Vigo, España.
}

\abstract{
Thermodynamic geometry of two-dimensional fluids has been investigated using a square-well model as a prototype fluid. A comparison with the three-dimensional case is performed in the subcritical and supercritical domains of thermodynamic space. In the subcritical region, it is found that the $R$-crossing method has a narrower range of validity for two-dimensional fluids compared to three-dimensional ones. On the other hand, in the supercritical region, an analysis of different Widom lines, including the $R$ Widom line, shows that for two-dimensional fluids these lines extend further into the supercritical region than their three-dimensional counterparts. A similar behavior is observed for the validity of the Clausius--Clapeyron equation in two-dimensional fluids.
}

\begin{document}

\maketitle

\section{Introduction}

The reduction of dimensionality often gives rise to novel physical phenomena absent in their three-dimensional (3D) counterparts, due, for instance, to enhanced fluctuations and modified interaction potentials \cite{Kosterlitz1978,Barber1980,Totsuji2004,Ruggeri2013}. The theoretical and experimental exploration of two-dimensional systems has been accelerated by the advent of atomically thin materials such as graphene \cite{Geim2009}, transition metal dichalcogenides \cite{Manzeli2017}, and artificial two-dimensional lattices in cold atom systems \cite{Liu2014}, all of which provide tunable platforms for studying fundamental physical processes in reduced dimensionality.

Unlike their 3D analogs, 2D fluids often exhibit distinct critical behavior, altered scaling relations, and different universality classes \cite{Kosterlitz1978,Barber1980}. Phenomena such as the Kosterlitz--Thouless transition, which has no direct analog in three dimensions, underscore the critical role that topological defects and fluctuations play in 2D phase transitions \cite{Kosterlitz1978}. Despite significant advances in understanding the thermodynamics of these systems, the thermodynamic landscape of 2D fluids, especially under extreme conditions such as supercritical regimes, remains only partially charted.
Several techniques have been used over the past decades to study the physics of low-dimensional systems, in particular for square-well fluids \cite{mishra1984,smit1991,singh1990,Smith1996,martinez2007,vortler2008,jimenez2008,rzysko2010,armas2013,gamez2020}. Among these techniques, thermodynamic geometry (TG) has emerged in recent years as a tool to explore thermodynamic properties and phase behavior in three-dimensional systems, but it has not yet been extensively applied to two-dimensional ones. An early formulation of TG was introduced by Weinhold \cite{Weinhold1975}, followed a few years later by alternative formulations proposed by Ruppeiner and Quevedo \cite{Ruppeiner1979,Quevedo2007}. 

In particular, the Ruppeiner approach, grounded in Riemannian geometry, interprets the second derivatives of the entropy as a metric structure on the thermodynamic state space.

In the case of two-dimensional systems, the fundamental principles of the theory remain unchanged. Its structure, based on the theory of thermodynamic fluctuations and on a Hessian metric, is preserved. What changes is the functional form of the thermodynamic potentials that define the metric structure. The consequences of this structural change have not yet been studied and undoubtedly deserve careful investigation. This theory not only provides insights into the nature of thermodynamic interactions, where the scalar curvature, $R$, is conjectured to reflect the correlation volume, but also encodes critical behavior and phase transitions in a coordinate-independent manner \cite{Ruppeiner1995}. For example, in several systems, $R$ diverges near second-order critical points, whereas changes in the sign of $R$ have been associated with a shift from attractive to repulsive intermolecular interactions \cite{Jaramillo2022}. This geometric language could be particularly appealing for complex systems such as supercritical fluids, where conventional thermodynamic indicators may be insufficient to characterize underlying microscopic changes.

Two-dimensional supercritical fluids (2D SCFs) may exhibit modified crossover behavior and distinct thermodynamic responses due to constrained phase space and altered fluctuation spectra \cite{Koch1983,Andrews1976,Jakli2006}. However, their thermodynamic characterization remains underdeveloped. Notably, the application of thermodynamic geometry to 2D SCFs offers a novel route to identify and classify supercritical crossovers and to quantify thermodynamic curvature in a low-dimensional setting where traditional tools may falter.\\

In this work, we investigate the thermodynamic geometry, within Ruppeiner's formalism, of two-dimensional fluids using a combination of statistical mechanical models and thermodynamic analysis, following the same methodologies that have been successfully applied previously to three-dimensional systems \cite{lopez2022,escamilla2024}. By studying the behavior of the Ruppeiner curvature across a range of thermodynamic conditions, including subcritical and supercritical ones, we explore its capacity to signal structural crossovers, critical remnants, and interaction regimes in reduced dimensions. This study not only extends the scope of thermodynamic geometry into new physical territory but also contributes to a deeper understanding of supercriticality in two-dimensional fluids, an emerging frontier with implications for both fundamental physics and technological applications.\\
This work is organized as follows. In Section~2, a brief summary of the basic aspects of thermodynamic geometry and the equations of state used in this work is presented. Section~3 is devoted to the main results, including a thermodynamic analysis of the subcritical and supercritical regions. Finally, the main conclusions are discussed in the last section.

\section{Theory}

\subsection{Thermodynamic geometry}

Within the framework of thermodynamic geometry developed by Ruppeiner \cite{Ruppeiner1979,Ruppeiner1995,Ruppeiner2012PRE}, the second derivatives of a thermodynamic potential are used to construct a metric associated with a thermodynamic system \cite{Ruppeiner2012PRE}, in a manner resembling the Einstein--Landau fluctuation theory, in which equilibrium fluctuations are obtained from the second-order expansion of the thermodynamic potential around equilibrium, leading to Gaussian statistics for extensive variables. The resulting mean-square fluctuations are governed by the curvature of the potential and are directly related to macroscopic response functions.

Several important features related to the thermodynamic scalar curvature $R$ have been studied in the literature \cite{May2012,May2013PRE,Ruppeiner2012PRE2,Jaramillo2019}. For example, its sign provides information about the nature of the effective interaction (attractive when $R<0$ and repulsive when $R>0$). The case $R=0$ indicates a null interaction, which is obtained for monocomponent ideal gases; however, this behavior is not observed in mixtures of ideal gases because the entropy includes a mixing contribution, which generates compositional correlations and fluctuations that are captured by Ruppeiner’s geometry, leading to a nonvanishing curvature of purely entropic origin \cite{Ruppeiner1995,Jaramillo2020}.\\

In the subcritical region, the $R$-crossing method allows one to reproduce the liquid--gas coexistence curve in the neighborhood of the critical point. This procedure consists of calculating the isotherms of the thermodynamic scalar curvature in both the liquid ($R_{l}$) and gas ($R_{g}$) phases and solving the equation $R_{l} = R_{g}$.
In addition, an important aspect related to this quantity is its anomalous behavior in the supercritical region, similar to that of response functions, whose extrema at fixed temperatures in the $(P,T)$ plane can be associated with the Widom line \cite{Speck2012,May2012,May2013PRE,Jaramillo2022}.

\subsection{Square-well systems in 3D and 2D}
In order to compare the thermodynamic geometry and supercritical behavior of 3D and 2D systems, we have selected the well-known square-well model. There is an extensive literature on this topic. A particularly interesting aspect relates the values of the critical parameters to the range of the potential. For example, the critical temperature varies monotonically with increasing potential range, in contrast to the critical density \cite{vega1992,orkoulas1999,pagan2005,scholl2005,dinpajooh2015,rzysko2012,reyes2013}.
The results presented in this work are obtained using equations of state derived from thermodynamic perturbation theory. For the 3D system, the Helmholtz free energy used is taken from Patel \textit{et al.} \cite{Patel2005} and is given by
\begin{equation}
    a_{_{3D}} = a^{Id}_{_{3D}} + a^{HS} + \beta a_{_{1}} + \beta^{2} a_{_{2}}, 
\end{equation} 
where $a^{Id}{{3D}}$ is the ideal contribution, $a^{HS}$ is the hard-sphere contribution, and the terms $\beta a_1$ and $\beta^2 a_2$ arise from perturbation theory, which involves the hard-sphere compressibility taken from \cite{Carnahan1969}. The validity of this equation of state is restricted to the range $1.2 \leq \lambda_{_{3D}} \leq 3.0$.

For two-dimensional systems, we employ the Helmholtz free energy formulation developed by Trejos \textit{et al.} \cite{Trejos2018}, which follows the same general functional form as in the three-dimensional case, although the specific terms differ.
\begin{equation}
    a_{_{2D}} = a^{Id}_{_{2D}} + a^{HD} + \beta a_{1} + \beta^{2} a_{2}, 
\end{equation} 
The validity range in this case lies within the interval $1.02 \leq \lambda_{_{2D}} \leq 12.0$. Note that this interval is much larger than that of the 3D square-well model.

To compare the extreme values of $R$ in the supercritical region, three response functions were selected: the heat capacity at constant pressure, the thermal expansion coefficient, and the speed of sound, which can be calculated in their reduced forms, respectively, as; $c^{*}_{p} \equiv \frac{c_{p}}{k} =  c^{*}_{v} + \frac{T^{*} \alpha_{p}^{*2}}{\rho^{*} \beta^{*}_{_T}},$ $\alpha^{*}_{p} \equiv \frac{\epsilon}{k} \alpha_{p} = \beta^{*}_{_T} \left( \frac{\partial P^{*}}{\partial T^{*}} \right)_V,$ and $c_{s}^{*2} \equiv \frac{m}{\epsilon} c_{s}^{2} = \frac{1}{\rho^{*} \beta^{*}_{s}},$
where $m$ is the mass of a single particle, $P^*=\epsilon P/B$ is the reduced pressure, $T^* = kT/\epsilon$ is the reduced temperature with $\epsilon$ the depth of the square well and $\rho^*=B\rho$ the reduced density where $B=\sigma^3$ or $B=\sigma^2$ for 3-D or 2-D systems, respectively. The reduced heat capacity at constant volume and the reduced adiabatic and isothermal compressibility are evaluated, respectively as; $c^{*}_v \equiv \frac{c_v}{k}= - T^{*}\frac{\partial^2 (T^{*} a) }{\partial T^{*2}},$ $\beta^{*}_{_T}  \equiv \frac{\epsilon}{B} \beta_{_T} = \frac{1}{\rho^{*} \frac{\partial P^{*}}{\partial \rho^{*}}}$ and $\beta^{*}_{s} \equiv \frac{\epsilon}{B} \beta_{_s}=  \beta^{*}_{_T} + \frac{T^{*} \alpha_{p}^{*2}}{\rho^{*} c^{*}_{p}}.$

\section{Results}

\begin{figure}
\centering
\includegraphics[width=0.9\columnwidth]{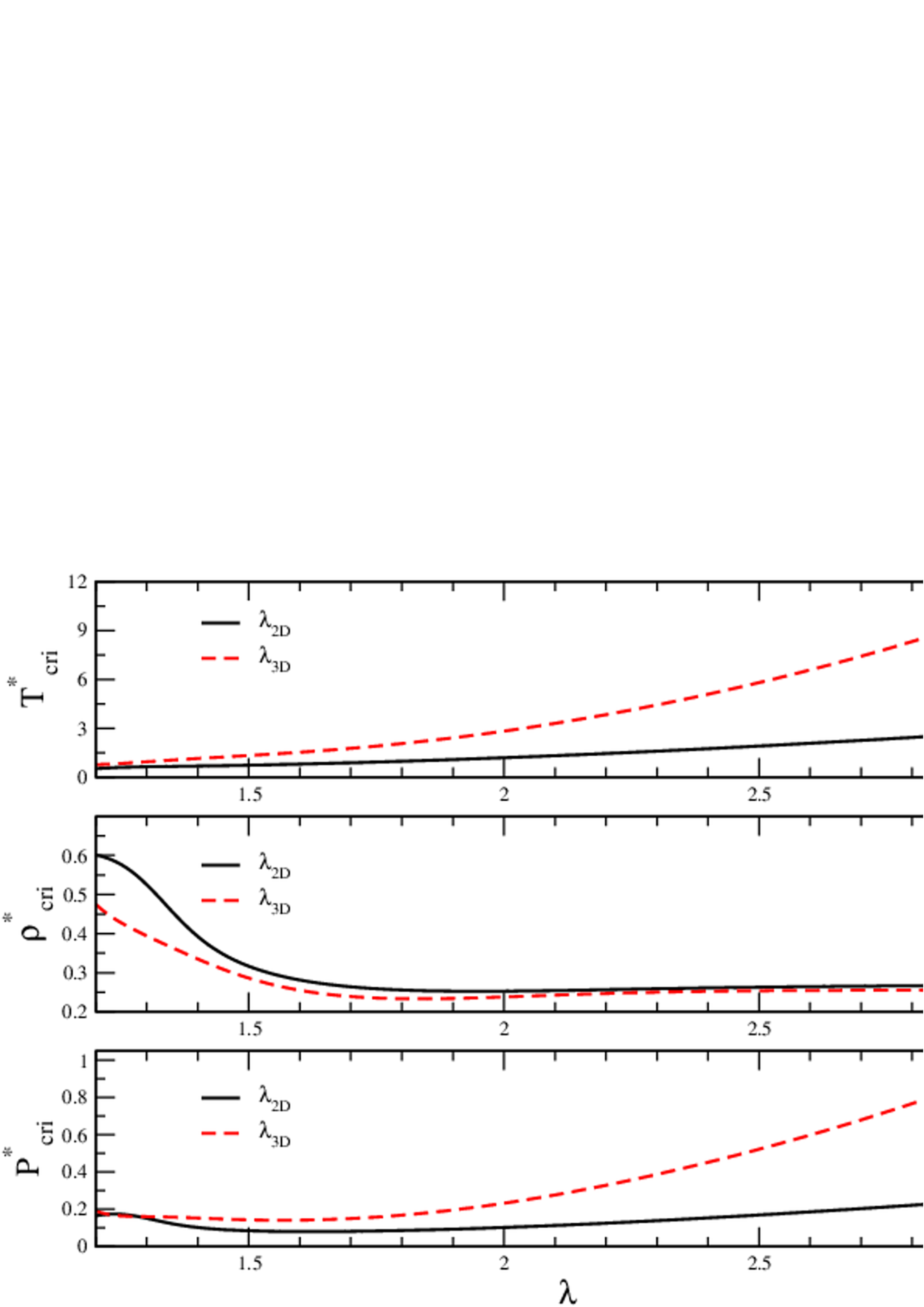}
\caption{Behavior of critical values as a function of the potential range for 2D and 3D square-well systems. From top to bottom: critical temperature, critical density, and critical pressure. } \label{CantCrit}
\end{figure}

In order to analyze the thermodynamic geometry behavior of 2D fluids and compare these results with 3D systems, four distinct ranges were chosen. Critical values for reduced temperatures, densities, and pressures associated with each $\lambda$ studied in this work are provided in Table \ref{CriticalValue}. Furthermore, Figure \ref{CantCrit} shows the behavior of critical quantities as a function of the potential's range. It can be observed that the critical temperature exhibits a monotonically increasing behavior with the range, while the critical pressure and density show a minimum when moving from short to long ranges. A reduction of the critical density with increasing dimensionality is also observed for short interaction ranges, as predicted by Mon and Percus \cite{mon1999}.

\begin{figure}
\centering
\includegraphics[width=0.9\columnwidth]{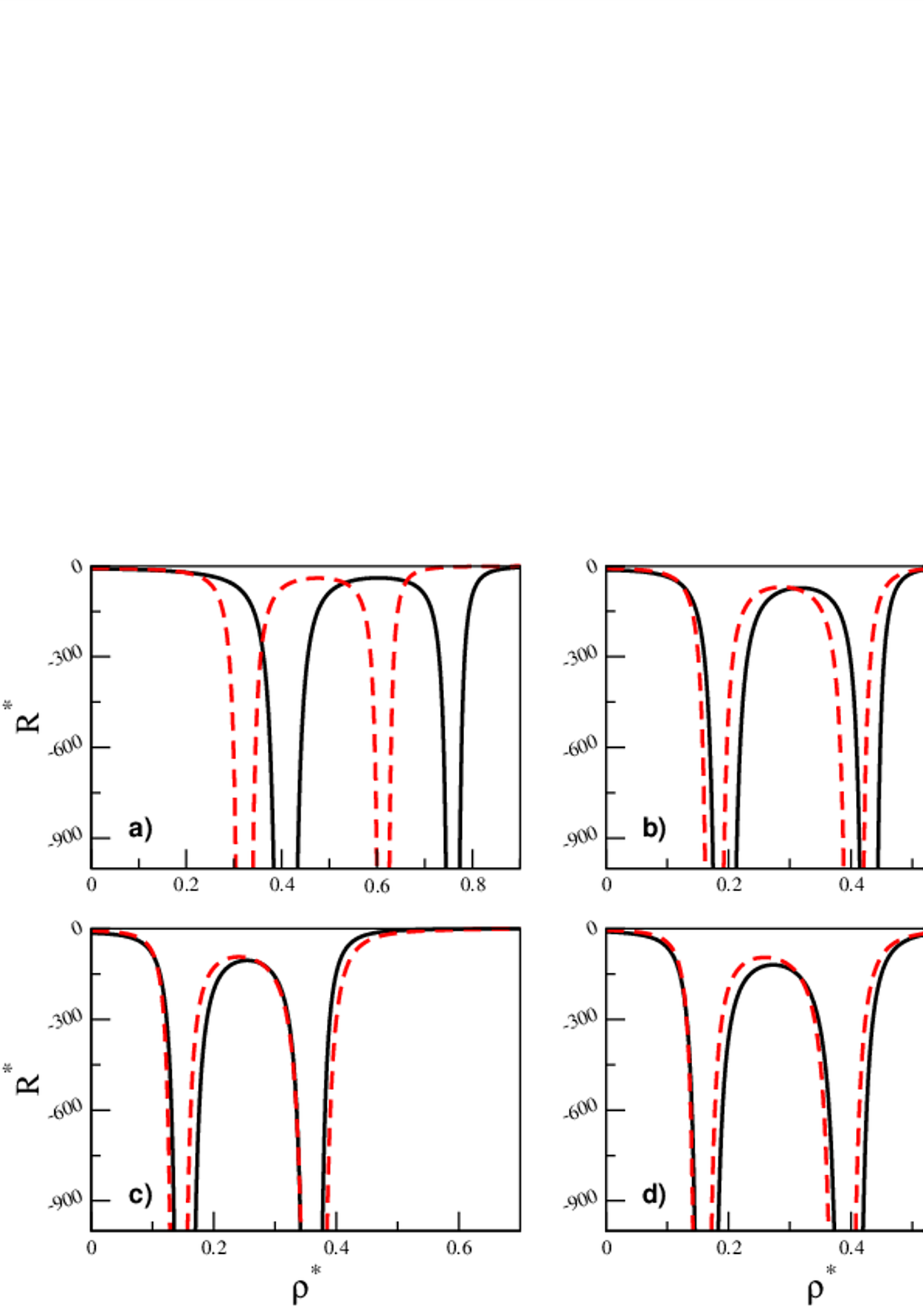}
\caption{Isotherms of the scalar curvature in the subcritical region at reduced temperature $T_{r} = 0.90$. In all plots, solid lines correspond to 2D square-well systems, and dashed red lines correspond to 3D square-well systems for the following ranges: a) $\lambda{{2D, 3D}} = 1.2$; b) $\lambda{{2D, 3D}} = 1.5$; c) $\lambda{{2D, 3D}} = 2.0$; d) $\lambda{_{2D, 3D}} = 3.0$.}\label{IsoSub}
\end{figure}

Figure \ref{IsoSub} presents the isotherms of the scalar curvature in the subcritical region for all the ranges considered in this study for 2D and 3D fluids. Divergences in the scalar curvature, which indicate the boundaries of the stability region for the fluids, are clearly observed. From these data, the complete spinodal curve can be extracted, since the positions of the divergences correspond exactly to points on this curve. Notably, the primary differences among these isotherms become evident at shorter ranges. For the larger \textit{equivalent} ranges, the curves are nearly indistinguishable. This behavior can be understood because, for these ranges, the system is near the mean-field limit, where a common scenario emerges. Despite these differences, the scalar curvature structure is the same for both 2D and 3D fluids, with only a shift in the positions of divergences observed when the dimension is reduced.\\

We have analyzed the behavior of $R$ in the vicinity of the critical point and found that, in all cases, the critical exponent is equal to $2$, corresponding to the expected mean-field value. This occurs because the equations of state employed are not able to capture the correct behavior in that region (Ising-like behavior).

\begin{figure}
\centering
\includegraphics[width=0.9\columnwidth]{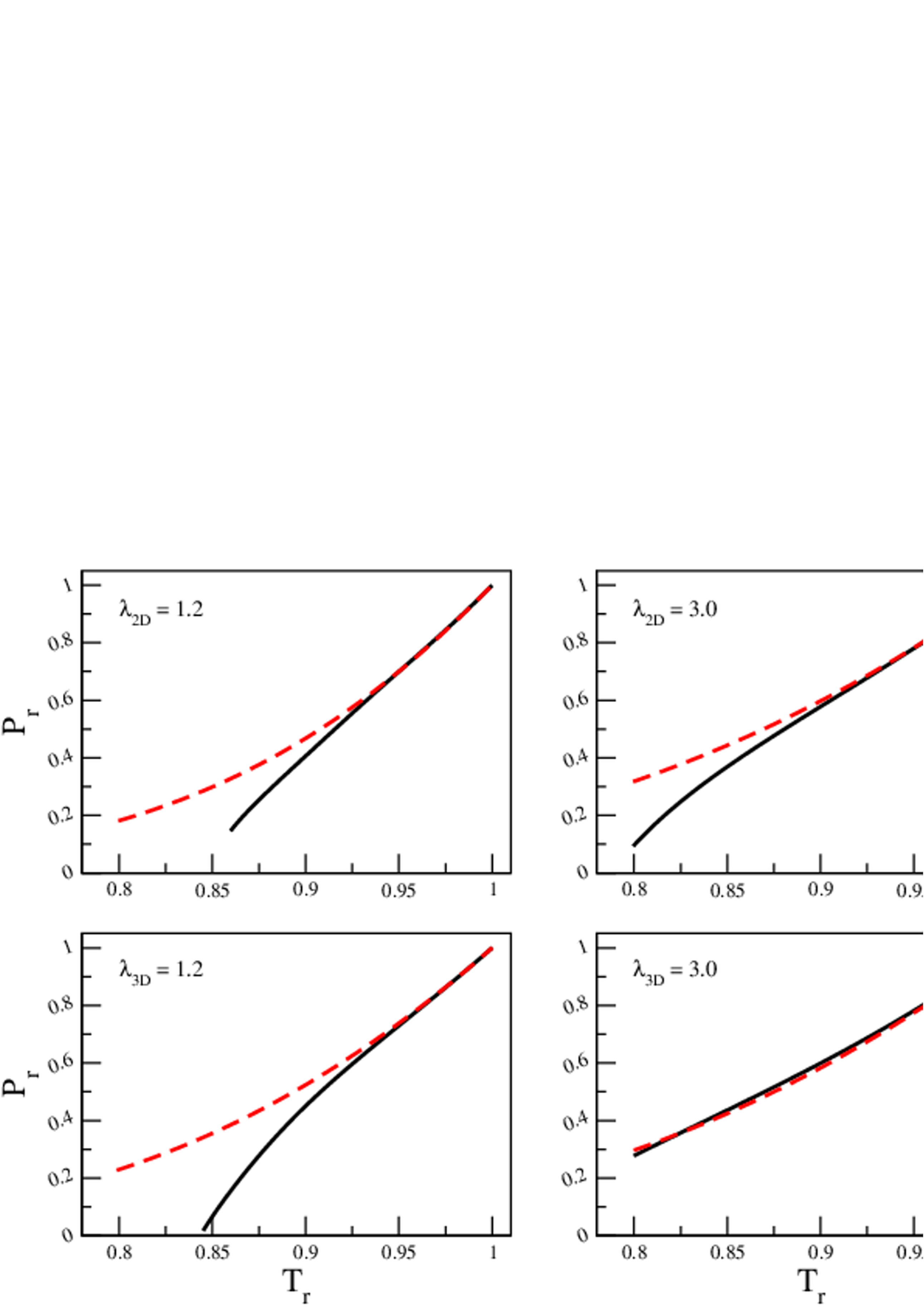}
\caption{Vapor–liquid coexistence line for 2D and 3D square-well systems. Solid black lines were calculated using the thermodynamic procedure, while dashed red lines were obtained using the R-crossing method.}\label{Rcros}
\end{figure}

\begin{figure}
\centering
\includegraphics[width=0.9\columnwidth]{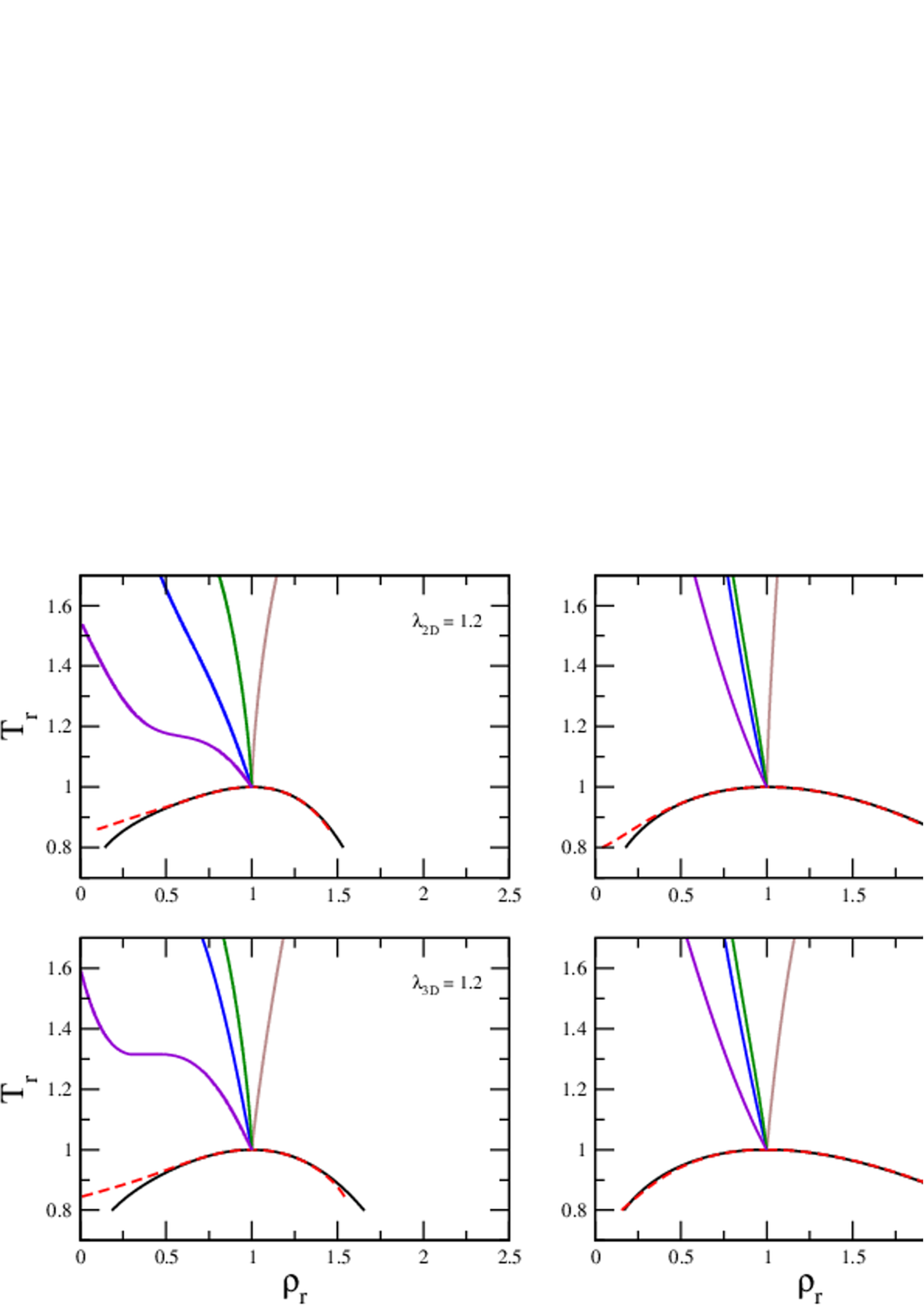}
\caption{Vapor–liquid coexistence line and extreme values of the response functions. Solid black lines were calculated using the thermodynamic procedure, while dashed red lines were obtained using the R-crossing method. For the extreme values: blue line, maximum of $\alpha_p$; brown line, maximum of $c_p$; green line, minimum of $R$; and violet line, minimum of the sound velocity.}\label{CoexWid}
\end{figure}

A key question regarding the differences between 2D and 3D systems within the framework of thermodynamic geometry is the validity of the R-crossing method. The R-crossing method exploits the feature of Ruppeiner’s thermodynamic geometry whereby the scalar curvature $R$ encodes the correlation volume of a phase. Phase coexistence is defined as the state points where $R$ is equal in the two phases, reflecting the equivalence of correlation volumes. In other words, the R-crossing method defines coexistence as the state points where the scalar curvature $R$ of the vapor and liquid phases are equal, as outlined in the Theory section. This provides a practical computational approach to determine phase equilibrium directly from thermodynamic geometry.
To address this, Figure \ref{Rcros} and \ref{CoexWid} presents the vapor-liquid coexistence lines, in the P-T and T-$\rho$ projections respectively, for both, the shorter and longer potential ranges considered in this study. Reduced units with respect to the critical ones were chosen to display the plots, i.e., $T_r = T/T_{{cri}}$ and $P_r = P/P_{{cri}}$. The solid black lines represent results obtained using the thermodynamic procedure, which involves solving for the equality of pressure and chemical potential, while the dashed red lines correspond to the R-crossing method. As observed, the geometrical method remains valid only in a small region near the critical point for systems with shorter potential ranges. As the potential range increases, the applicability of the R-crossing method extends, with its validity generally larger for 3D fluids compared to 2D ones. Both 2D and 3D systems exhibit asymptotic behavior at longer potential ranges (mean-field limit). Using a 5\% deviation criterion relative to the thermodynamic procedure, it is found that the R-crossing method fails beyond this threshold, approximately 20\% below the critical point for 3D fluids and 14\% for 2D systems. Although the 5\% criterion is arbitrary, it allows for a meaningful comparison, and the plots show deviations larger than this percentage. Clearly, if this threshold is reduced, the region of agreement also decreases; however, the relevant point is the comparative analysis itself.\\

Since the R-crossing method is based on Ruppeiner's conjecture, this conjecture appears to break down more quickly for 2D systems than for 3D ones. One might tentatively attribute this result to the different critical behavior of thermodynamic quantities in 2D and 3D systems. However, since in this work we are using theoretical equations of state that do not include finite-size effects around the critical region, both equations for 2D and 3D SW fluids exhibit mean-field behavior in the vicinity of the critical point, independent of dimensionality. Therefore, this is not a valid explanation for the observed dimensional dependence. Further research is needed to obtain valid arguments explaining these differences.\\

\begin{figure}
\centering
\includegraphics[width=0.9\columnwidth]{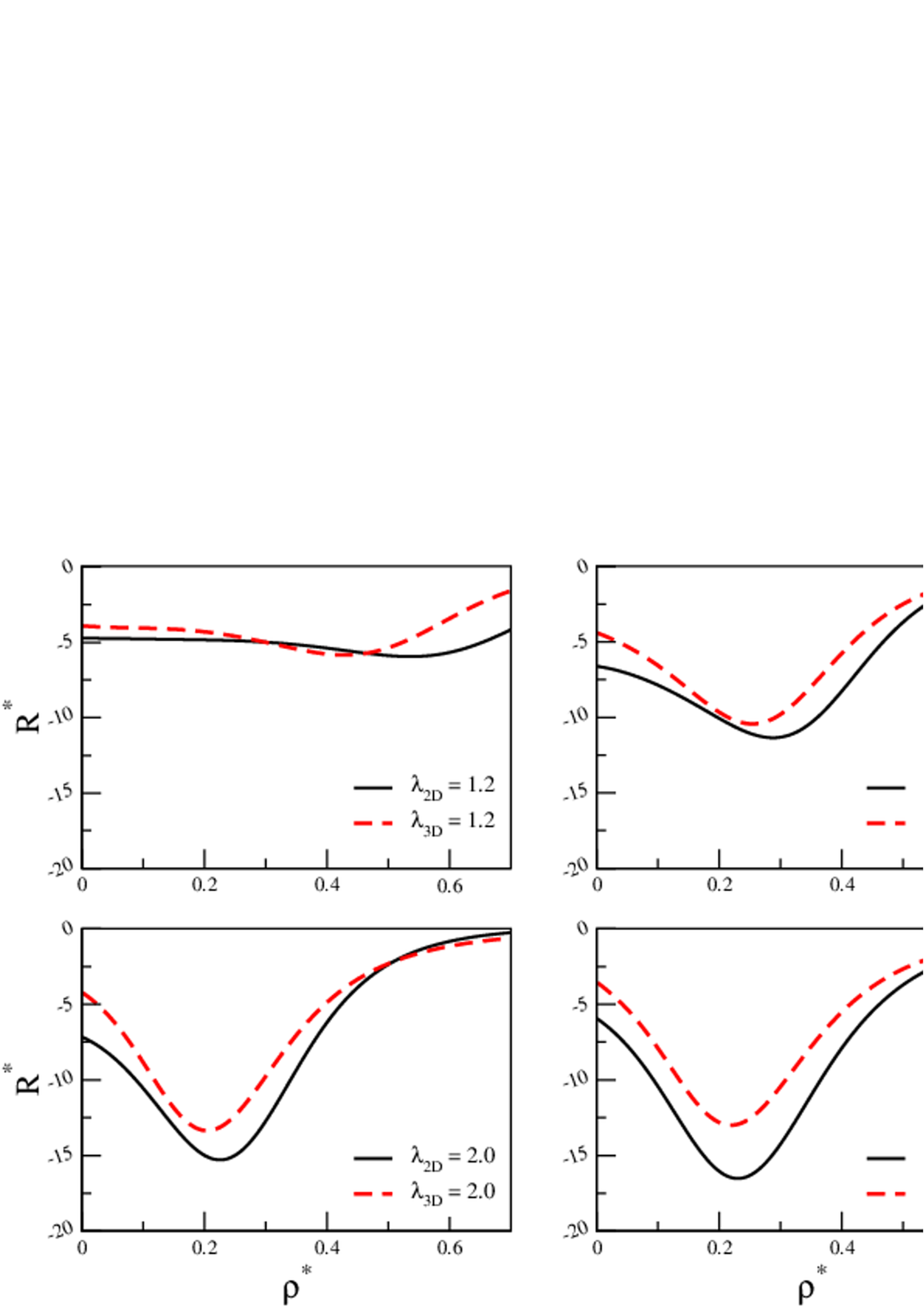}
\caption{Isotherms of the scalar curvature in the supercritical region for several ranges at reduced temperature $T_{_r} = 1.50$.}\label{IsoSup}
\end{figure}

In the supercritical region, the isotherms of the scalar curvature are shown in Figure \ref{IsoSup}. As expected, no divergences are observed in this region. However, typical anomalies near the critical region are evident, with their magnitude being more pronounced for larger ranges compared to shorter ones. Similar to the subcritical region, the differences among the isotherms of the scalar curvature for 2D and 3D systems are most noticeable at shorter ranges, while a common limiting behavior (the position of the minimum) is observed for the longer ranges. Both in the subcritical and supercritical regions, there is a change in the sign of R at high densities, which is often associated with the approach to the solid phase. However, this occurs at densities that exceed the validity range of the equations of state used in this work.\\

\begin{figure}
\centering
\includegraphics[width=0.9\columnwidth]{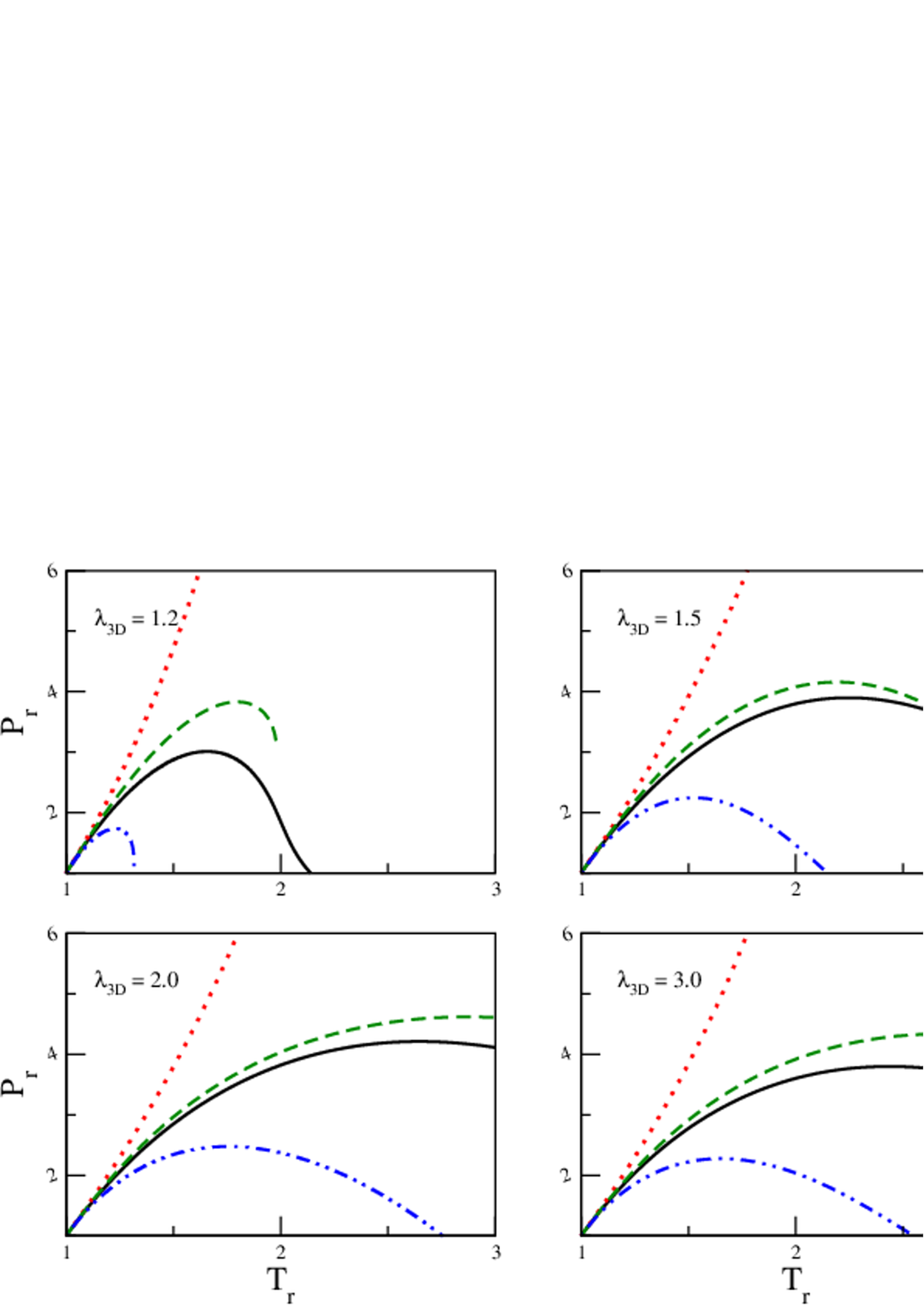}
\caption{Extremes of the response functions for 3D square-well systems. Solid black line, maximum of $\alpha_p$; dotted red line, maximum of $c_p$; dashed green line, minimum of $R$; and dot-dashed blue line, minimum of the sound velocity. }\label{3Dwidom}
\end{figure}

\begin{figure}
\centering
\includegraphics[width=0.9\columnwidth]{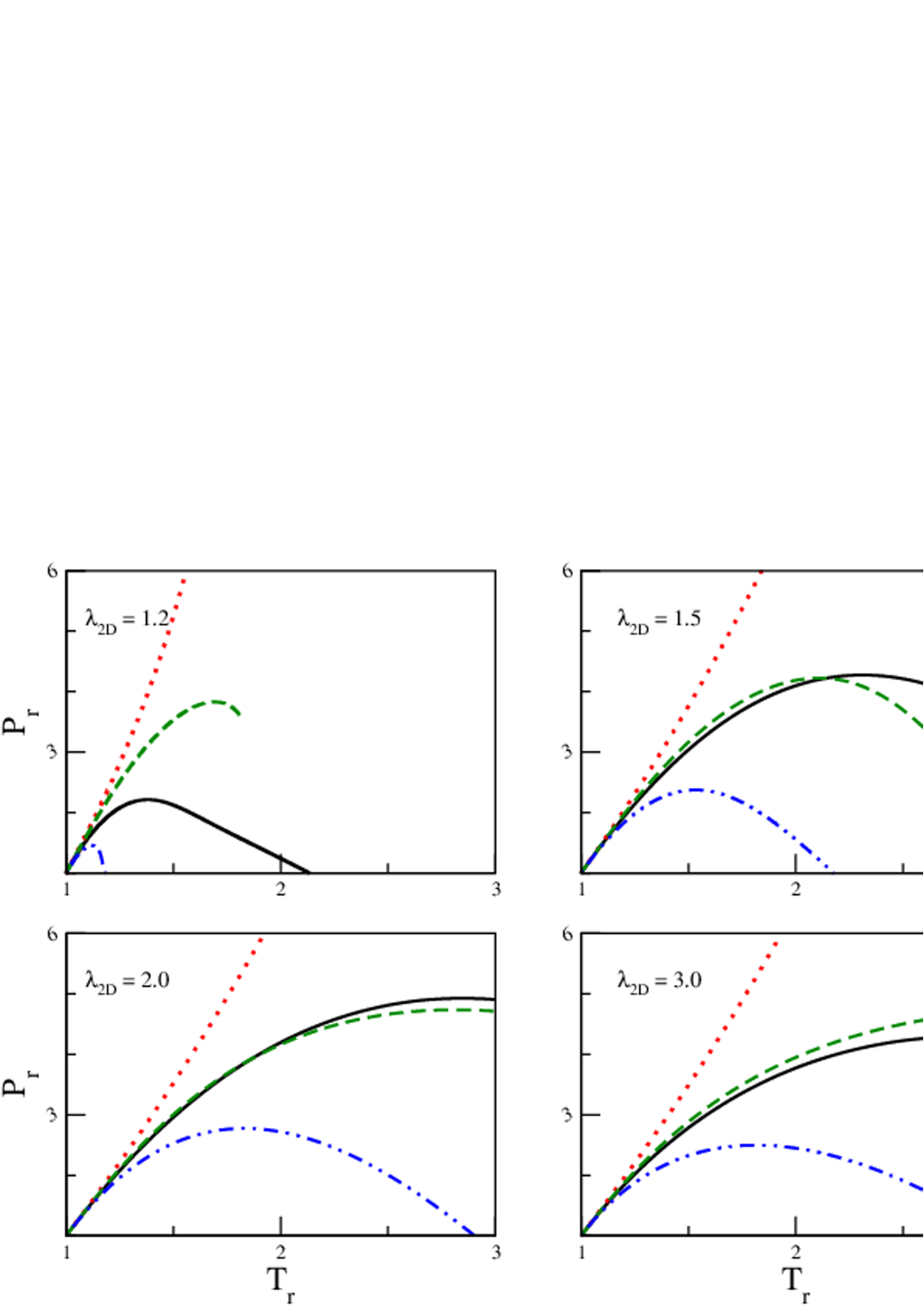}
\caption{Extremes of the response functions for 2D square-well systems. Solid black line, maximum of $\alpha_p$; dotted red line, maximum of $c_p$; dashed green line, minimum of $R$; and dot-dashed blue line, minimum of $c_s$.}\label{2Dwidom}
\end{figure}

One of the most interesting behaviors observed in the supercritical region is the appearance of separatrix lines. Among the several proposed lines, Widom lines are perhaps the most studied. Figures \ref{3Dwidom} and \ref{2Dwidom} show the Widom lines associated with the heat capacity (at constant pressure) $c_p$, the expansion coefficient $\alpha_p$, the sound velocity $c_s$, and the thermodynamic scalar curvature $R$, for 3D and 2D square-well fluids in the T-P projection. In Figure \ref{CoexWid} this lines are plotted in the T-$\rho$ plane for both the
shorter and longer potential ranges considered in this
study. As usual, the $c_p$ and $\alpha_p$ Widom lines are obtained from the maxima of these response functions in the supercritical region, whereas the $R$ and $c_s$ Widom lines are obtained from the minima of $R$ and $c_s$ in the same region.\\

As can be seen, the qualitative behavior is the same for all considered ranges, independent of dimensionality. A non-monotonic behavior of the Widom lines corresponding to $\alpha_p$, $c_s$, and $R$ is observed in all cases, contrasting with the monotonic behavior of the $c_p$ Widom line. This behavior is common to both 2D and 3D systems; however, an interesting difference emerges: the Widom lines for 2D systems extend further than those for 3D systems. To the best of our knowledge, this is the first time this observation has been reported in the literature. Since supercritical fluids are used in several industrial processes, and crossing the Widom line implies that a small change in temperature or pressure results in large changes (compared to other supercritical regions) in the corresponding thermodynamic properties, this observation opens an interesting window for potential applications of 2D systems under supercritical conditions.

\begin{figure}
\centering
\includegraphics[width=0.9\columnwidth]{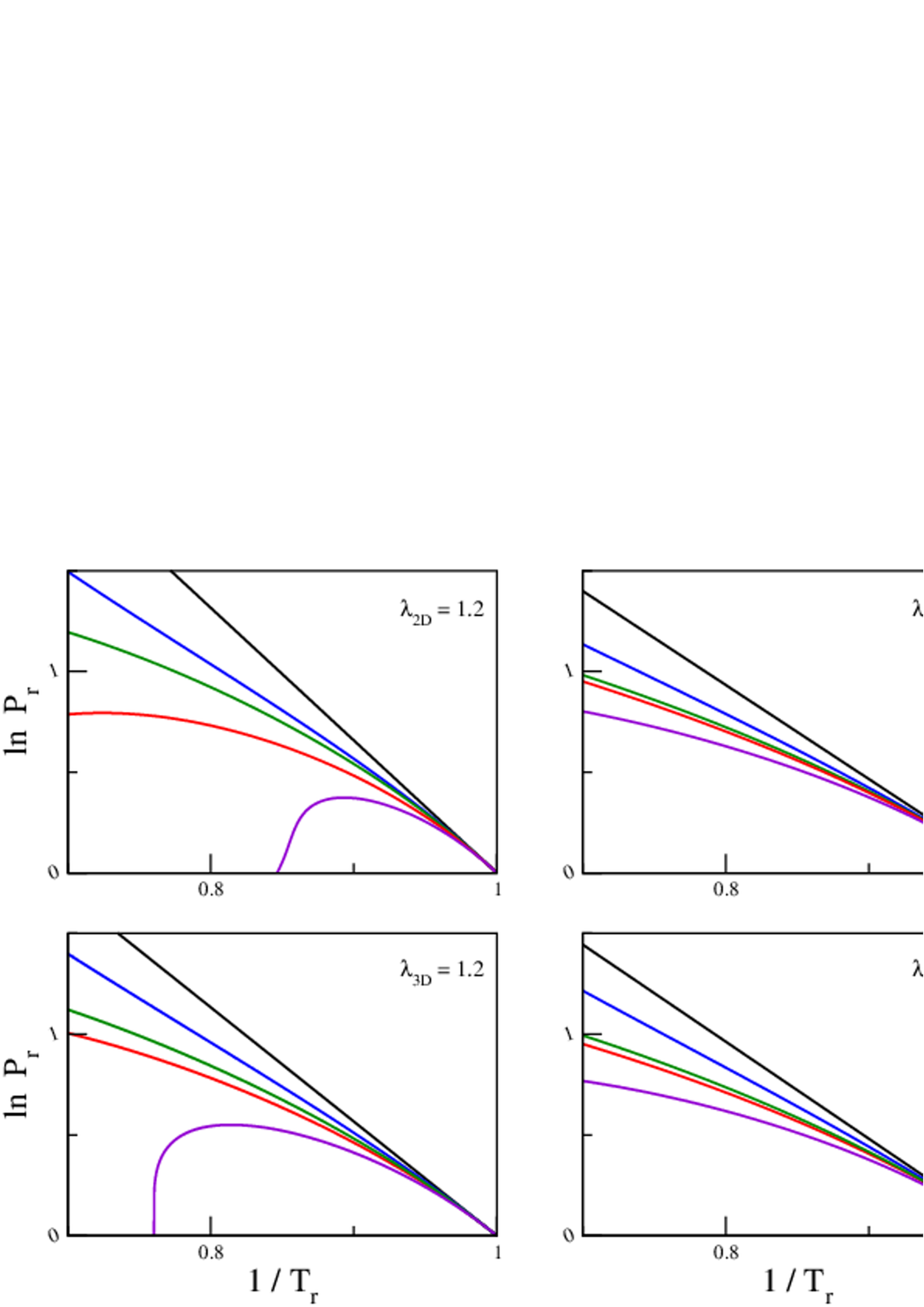}
\caption{Natural logarithm of the pressure associated with the extreme values of the response functions. From top to bottom: global fit of the response functions in the neighborhood of the critical point; maximum of $C_{P}$; minimum of $R$; maximum of $\alpha_p$; and minimum of $c_s$. }\label{Clausius}
\end{figure}

The extreme values of the four thermodynamic quantities associated with the Widom lines were used to fit the Clausius–Clapeyron equation in a temperature neighborhood of the critical point. This relation can be expressed as

\begin{equation}
    \ln P_r = A_0\left( \frac{1}{T_r} - 1 \right), \label{Clausequ}
\end{equation}

when working with reduced units relative to the critical ones, where $A_0$ is a constant that can be fitted. An arbitrary 0.05\% temperature neighborhood was used to perform the fits, within which 20 equally spaced data points were taken from each of the response functions and the scalar curvature. The values of $A_0$ obtained are listed in Table \ref{CriticalValue} for all ranges considered in this study. In Figure \ref{Clausius}, these fits are plotted on a logarithmic scale (for pressure) with the extreme values of the response functions studied here and of $R$, as a function of inverse temperature, for both the shorter and larger ranges considered. In all cases, the values associated with $c_p$ remain in better agreement with the fit of Eq. \ref{Clausequ} compared to the other thermodynamic quantities, with the sound velocity being the response function that deviates most rapidly. An interesting result is that the Clausius–Clapeyron equation remains valid over a larger neighborhood around the critical point for 2D systems compared with 3D ones. This is the opposite behavior observed for the R-crossing method in the subcritical region.

\begin{table*}[!t]
\centering
\resizebox{\textwidth}{!}{%
\begin{tabular}{cccccccccc}
\hline
$\lambda_{_{2D}}$ & $T^*_{_{cri}}$ & $\rho^*_{_{cri}}$ & $P^*_{_{cri}}$ & $A_0$ & $\lambda_{_{3D}}$ & $T^*_{_{cri}}$ & $\rho^*_{_{cri}}$ & $P^*_{_{cri}}$ & $A_0$ \\ 
\hline
$1.2$ & $0.521152$ & $0.600969$ & $0.165008$ & $-6.57799$ & $1.2$ & $0.762291$ & $0.474953$ & $0.192509$ & $-5.66462$ \\ 
$1.5$ & $0.732469$ & $0.316642$ & $0.084924$ & $-5.01870$ & $1.5$ & $1.329069$ & $0.286359$ & $0.143413$ & $-5.05733$ \\ 
$2.0$ & $1.19178$ & $0.252846$ & $0.102262$ & $-4.70906$ & $2.0$ & $2.824792$ & $0.232517$ & $1.329069$ & $-4.89200$ \\ 
$3.0$ & $2.826349$ & $0.267832$ & $0.260467$ & $-4.66484$ & $3.0$ & $10.32879$ & $0.255257$ & $0.960587$ & $-4.81516$ \\ 
\hline
\end{tabular}%
}
\caption{Critical values for 2D and 3D square-well systems for all ranges considered in this work. The values of $A_0$ are the fitted parameters obtained for the Clausius–Clapeyron equation in Eq. \ref{Clausequ}.}
\label{CriticalValue}
\end{table*}

\section{Conclusions}
We have studied the thermodynamic geometry of two-dimensional fluids using an expression for the Helmholtz free energy derived from thermodynamic perturbation theory. Given its versatility and well-known thermodynamic properties, a square-well fluid was selected for this study to investigate possible differences between the thermodynamic geometry of two- and three-dimensional fluids.
Both subcritical and supercritical regions of the thermodynamic space were analyzed. Isotherms of the scalar curvature for two- and three-dimensional square-well fluids, for different ranges, were obtained and plotted as a function of density. The same qualitative behavior is observed, showing the typical divergences along the spinodal curve in the subcritical region and anomalies near the critical point in the supercritical region. However, when the R-crossing method was implemented, a difference was found: the method loses its validity more quickly for two-dimensional fluids. The R-crossing method relies on the equality of the correlation length between both phases, and since the correlation length is shorter for two-dimensional fluids compared to three-dimensional ones, this may explain the observed behavior. At present, this is not completely clear, and further studies are needed to clarify this issue.\\
In the supercritical region, in addition to the scalar curvature, several Widom lines were obtained, defined by the extrema of the isobaric heat capacity, expansion coefficient, sound velocity, and scalar curvature. The same qualitative behavior is observed for both two- and three-dimensional cases across all ranges studied.   Our results show that in two-dimensional fluids, the Widom lines associated with the extrema of response functions extend  deeper into the supercritical region than in three-dimensional fluids, reflecting the enhanced fluctuations inherent to reduced dimensionality.\\
The validity of the Clausius–Clapeyron equation in the supercritical region was also examined, showing that this relation is satisfied in a region further from the critical point for two-dimensional systems.\\
Thus, although several thermodynamic geometric behaviors are qualitatively similar for two- and three-dimensional fluids, the larger fluctuations associated with two-dimensional systems induce some important quantitative differences.
Future work could refine these results by employing equations of state that more accurately capture the fluid’s behavior near the critical point. Furthermore, while square-well potentials are discontinuous, applying this formalism to continuous potentials, such as the Lennard-Jones potential \cite{Homes2025}, would certainly be of great interest. These topics remain promising avenues for future work.

\section{Acknowledgments}
One of us (J.T.A.) would like to thank the Departamento de F'isica Aplicada of the Universidade de Vigo for the facilities provided and the warm welcome during a sabbatical stay, during which part of this work was carried out. We also acknowledge the financial support from the Universidad de Guanajuato and SECIHTI (through SNII), which made this stay possible.

\bibliographystyle{unsrt}
\bibliography{biblio}

\end{document}